\documentclass[usegraphicx]{mn2e}

\usepackage{graphicx}
\usepackage{afterpage}

\newcommand\Msun{{\,M_\odot}}

\title[]{The effect of gas fraction on the morphology and time-scales of disc galaxy mergers}

\author[J.M. Lotz et al.]{Jennifer M. Lotz,$^1$\thanks{NOAO Leo Goldberg Fellow},  
Patrik Jonsson,$^2$ T.J. Cox,$^3$\thanks{W.M. Keck Fellow} and Joel R. Primack$^2$ \\
$^1$ National Optical Astronomical Observatory, 950 N.  Cherry Avenue, Tucson, AZ 85719 USA; lotz@noao.edu \\
$^2$ Department of Physics, University of California, Santa Cruz, CA, 95064 USA \\
$^3$ Harvard-Smithsonian Center for Astrophysics, 60 Garden St.,  Cambridge, MA, 02138 USA}

\begin{document}

\date{submitted 17 August 2009; resubmitted 8 December 2009}

\pagerange{\pageref{firstpage}--\pageref{lastpage}} \pubyear{2009}

\maketitle

\label{firstpage}

\begin{abstract}
Gas-rich galaxy mergers are more easily identified by their disturbed morphologies than mergers with less gas.
Because the typical gas fraction of galaxy mergers is expected to increase with redshift,  
the under-counting of low gas-fraction mergers may bias morphological estimates of the evolution of 
galaxy merger rate.  To understand the magnitude of this bias,  we explore the effect of gas fraction on 
the morphologies of a series of simulated disc galaxy mergers.  With the resulting $g$-band images,
we determine how the time-scale for identifying major and minor galaxy mergers via close projected pairs and 
quantitative morphology  (the Gini coefficient $G$,   the second-order moment of the brightest 20\% of the 
light $M_{20}$, and asymmetry $A$) depends on baryonic gas fraction $f_{gas}$.  Strong asymmetries
last significantly longer in high gas-fraction mergers of all mass ratios, with time-scales ranging from 
$\leq 300$ Myr for $f_{gas} \sim $ 20\%  to $\geq$ 1 Gyr for $f_{gas} \sim $ 50\%.  Therefore the strong 
evolution with redshift observed in the fraction of asymmetric galaxies may reflect evolution in the gas properties 
of galaxies rather than the global galaxy merger rate. On the other hand, the time-scale for identifying 
a galaxy merger via $G-M_{20}$ is weakly dependent on gas-fraction ($\sim$ 200-400 Myr), consistent with the weak evolution observed for $G-M_{20}$ mergers. 
\end{abstract}


\section{INTRODUCTION}
Gas plays an important role in shaping the properties of galaxy mergers and their remnants.  Unlike stars and dark matter, gas can cool radiatively
and therefore lose kinetic energy efficiently.   Cold gas is converted into stars,  which in turn produce dust and supernovae.   The varying 
gravitational fields of galaxy mergers can draw gas out into long tidal tails that eventually form tidal dwarf galaxies (e.g. Barnes \& Hernquist 1996; Bournaud \& Duc 2006;
Wetzstein et al. 2007)  
or funnel gas into the galaxies' centers to spark  strong central starbursts (e.g. Mihos \& Hernquist 1996; Martig \& Bournaud 2008; di Matteo et al. 2008) 
and feed active galactic nuclei (e.g. Di Matteo, Springel, \& Hernquist 2005; Hopkins et al. 2005, 2008; Johansson et al. 2009;  de Buhr et al. 2009).  
Low density gas shocked by mergers can linger as hot X-ray emitting gaseous haloes (e.g. Cox et al. 2004, 2006a; Sinha \& Holley-Bockelmann 2009).   
The spheroidal remnants of gas-rich mergers can reform small discs out of the remaining gas (Barnes 2002; Cox et al. 2006b; Naab, Jesseit, \& Burkert 2006;
 Khalatyan et al. 2008),  
while the remnants of extremely gas-rich mergers can preserve discs large enough to resemble present-day spiral galaxies (e.g. Springel \& Hernquist 2005; 
Robertson et al. 2006b; Hopkins et al. 2009).  The kinematic and structural properties of gas-poor mergers, on the other hand, are
less rotationally supported and less centrally concentrated,  similar to present-day massive elliptical galaxies (e.g.  Boylan-Kolchin, Ma, \& Quataert 2005, 2006; 
Cox et al. 2006c; Robertson et al. 2006a; Naab, Khochfar, \& Burkert 2006)

In order to understand the formation of today's galaxies, astronomers need to measure the frequency of gas-rich and gas-poor mergers as
well as the global galaxy merger rate.   At present,  there are few direct measurements of cold gas in distant galaxies. Rest-frame optical
colours are often used as a proxy for gas content  with the assumption that blue star-forming galaxies are gas-rich and red quiescent galaxies are gas-poor.
Massive red galaxies were 2$-$4 times less common at $z\sim1$ than today (e.g. Bell et al. 2004; Faber et al. 2007; Brown et al. 2007), 
suggesting that gas-poor galaxies were rare at early times.    Several recent studies of the evolution of close pairs of galaxies have detected strong evolution
in the colours of pre-merger systems. In a study of kinematic pairs of galaxies in the Extended Groth Strip,  
Lin et al. (2008) found that kinematic pairs are three times more likely to both be blue  and four times less likely to 
both be red at $z \sim 1$ than in the local universe. 
Similar studies in the GOODS and COSMOS fields have also found strong evolution in red pairs out to $z \sim 1$ 
(Bundy et al. 2009; de Ravel et al. 2008).

\begin{table*}
  \centering
  \begin{minipage}{168mm}
    \caption{Initial Galaxy Conditions}
    \begin{tabular}{@{}lccrllllccccr@{}} 
      \hline
       Model  &  $N_{part}$\footnote{Number of particles in total, dark matter, gas, stellar disc, and stellar bulge for GADGET simulations} & $M_{vir}$\footnote{Virial mass} & $C$\footnote{Dark matter halo concentration}  & $M_{bary}$\footnote{Baryonic mass} & $M^*_{disc}$\footnote{Mass of stellar disc} & $M^*_{b}$\footnote{Mass of stellar bulge} & $M_{gas}$\footnote{Mass of gaseous disc} & $f_{b}$\footnote{Fraction of baryons in the bulge} & $f_{gas}$\footnote{Fraction of baryons in gas} & $R_{disc}$\footnote{Scale-length of stellar disc}  & $R_{b}$\footnote{Scale-length of bulge}  &$R_{gas}$\footnote{Scale-length of gaseous disc} \\
     &  & ($\Msun$) &  &($\Msun$) &($\Msun$) & ($\Msun$) & ($\Msun$)  &  &   & (kpc)  & (kpc)  & (kpc) \\
     \hline
 G3   & $2.4, [1.2, 0.5, 0.5, 0.2]\cdot10^5$  & $1.2\cdot10^{12}$  &6   & $6.2\cdot10^{10}$  &$4.1\cdot10^{10}$  &$8.9\cdot10^9$   &$1.2\cdot10^{10}$ & 0.14 &0.19  &2.85  & 0.62   &  8.55 \\
 G2   & $1.5,  [0.8, 0.3, 0.3, 0.1] \cdot10^5$  & $5.1\cdot10^{11}$  &9   & $2.0\cdot10^{10}$  &$1.4\cdot10^{10}$  &$1.5\cdot10^9$   &$4.8\cdot10^{9}$  & 0.08 &0.24  &1.91  & 0.43   &  5.73 \\
 G1   & $9.5, [5.0, 2.0, 2.0, 0.5]\cdot10^4$  & $2.0\cdot10^{11}$  &12  & $7.0\cdot10^{9}$   &$4.7\cdot10^{9}$   &$3.0\cdot10^8$   &$2.0\cdot10^{9}$  & 0.04 &0.29  &1.48  & 0.33   &  4.44 \\
\hline
 G3gf1 &$2.4, [1.2, 0.5, 0.5, 0.2]\cdot10^5$  & $1.2\cdot10^{12}$  &6  &  $6.2\cdot10^{10}$  &$2.9\cdot10^{10}$  &$8.9\cdot10^9$   & $2.4\cdot10^{10}$ & 0.14 & 0.39  &2.85  & 0.62   &  8.55 \\
 G3gf2  &$2.4, [1.2, 0.5, 0.5, 0.2]\cdot10^5$ & $1.2\cdot10^{12}$  &6  &  $6.2\cdot10^{10}$  &$2.0\cdot10^{10}$  &$8.9\cdot10^9$   & $3.3\cdot10^{10}$ & 0.14 & 0.53  &2.85  & 0.62   &  8.55 \\
 G3gf2a  &$3.1, [1.2, 1.4, 0.3, 0.2]\cdot10^5$ & $1.2\cdot10^{12}$  &6  &  $6.2\cdot10^{10}$  &$2.0\cdot10^{10}$  &$8.9\cdot10^9$   & $3.3\cdot10^{10}$ & 0.14 & 0.53  &2.85  & 0.62   &  8.55 \\
\hline
 Sbc  & $1.7, [1.0, 0.3, 0.3, 0.1] \cdot10^5$  & $8.1\cdot10^{11}$  &11  & $1.0\cdot10^{11}$  &$3.9\cdot10^{10}$  &$9.7\cdot10^9$     &$5.3\cdot10^{10}$ & 0.10 &0.52  &5.50  & 0.45   & 16.50 \\
\hline
\end{tabular}\label{gparstab}
\end{minipage} 
\end{table*}

\begin{figure*}
\includegraphics[width=168mm]{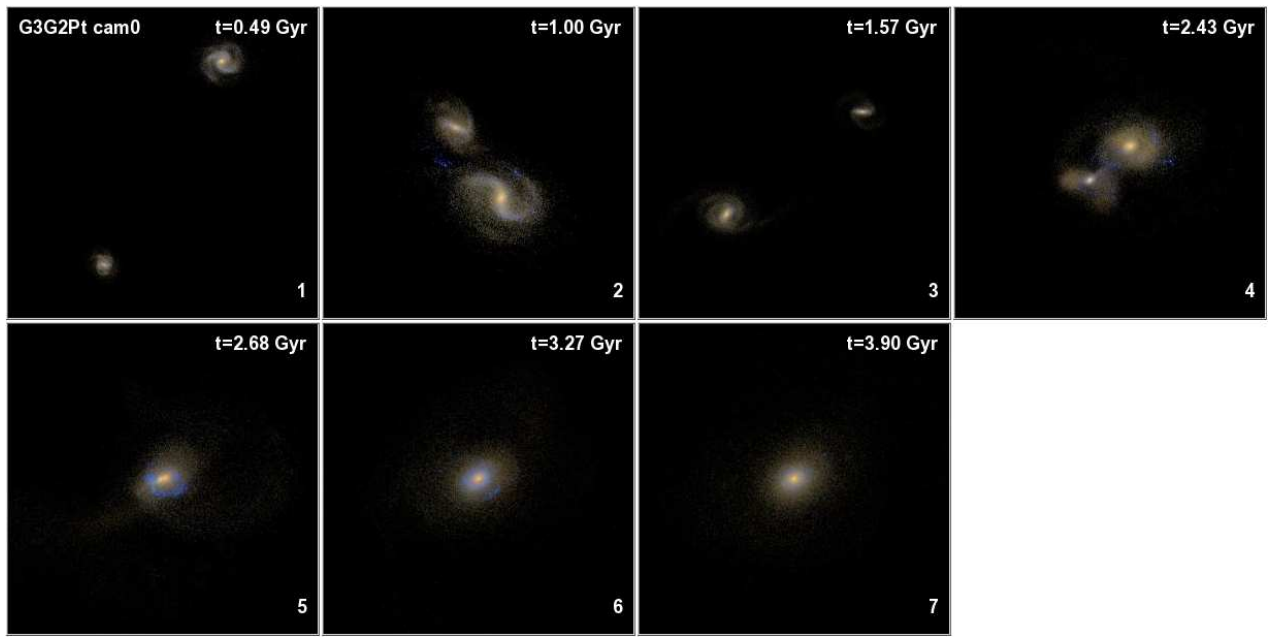}
\caption{ $u-r-z$ composite colour images  including dust extinction
for the  G3-G2 simulation with  $f_{gas}$ = 19\%  and 3:1 baryonic mass ratio.  The G3 primary galaxy is 
viewed face-on (camera 0).  The images show  the initial galaxies [1], the first pass [2], the maximal separation after the first pass [3], the second
pass [4],  the final merger [5], the post-merger [6], and the remnant $\sim$ 1 Gyr after the merger [7]. 
The field of view for panels 1 and 2 is 200 kpc, while the field of view for the other images is 100 kpc.   } \label{gf0col}
\end{figure*}

Morphological disturbances are an independent and efficient way to identify on-going galaxy mergers. 
The majority of $z \leq 1$ merger candidates with strong visual disturbances, asymmetries or double nuclei appear blue (e.g. Bell et al. 2005; 
Lotz et al. 2008a;  Jogee et al. 2009).  
But gas-rich mergers exhibit strong disturbances for longer periods of time than gas-poor mergers (Bell et al. 2006), and
therefore are more likely to be found.  Very low surface brightness tails and shells may persist in gas-poor mergers for $>$1 Gyr (Van Dokkum 2005), 
but these features are not easily detected at high redshifts.  Given that the strong morphological signatures of
purely dissipationless mergers last for less time than the signatures of a moderately gas-rich merger,  morphological estimates of the galaxy merger fraction
may under-count gas-poor galaxy mergers.  

The effect of gas fraction on the morphologies of disc-disc galaxy mergers has not yet been fully explored. 
Most theoretical work has focused on the impact of gas on the structure and kinematics of merger remnants (e.g. Boylan-Kolchin et al. 2005;
Cox et al. 2006b; Robertson et al. 2006a,b; Hopkins et al. 2009).  Only a few studies have examined the properties of on-going mergers with different gas properties 
(Bournaud et al. 2005; Lotz et al. 2008b). It remains unknown if large scale asymmetries and other disturbances scale with gas fraction,  or are equally evident in all 
mergers above a critical gas threshold. In order to understand the biases associated with detecting mergers of varying gas fractions, this work explores the effect of gas fraction on 
the time-scales for identifying simulated major and minor mergers via different quantitative methods ($G-M_{20}$, asymmetry, close pairs).

This paper continues the work first presented in Lotz et al. 2008b (hereafter Paper 1), and Lotz et al. 2009a (Paper 2).  
Because it is impossible to derive the initial merger parameters (mass ratio, gas fraction, orbits, etc.) 
from observations of on-going  highly disturbed mergers,  our approach is to simulate mergers with a wide range in parameter space and 
determine which parameters are important for the strength and duration of morphological disturbances.  
In Paper 1, we studied a series of  equal-mass disc merger simulations.
We found that the time-scales for identifying a particular galaxy merger could be quite different depending on the gas properties of the merging galaxies 
as well as the method used to find the merger.  However,  because the high gas-fraction and moderate gas-fraction galaxy models (Sbc and G3) presented in Paper 1  also
had different dark matter concentrations and gas disc scale-lengths,  it was not possible to determine if the baryonic gas fraction was the
governing factor.  Paper 2 explored the effect of mass ratio and orbital parameters for a series of unequal-mass disc mergers with gas fractions
fixed at local values.  We found that different morphological indicators probed different ranges of mass ratio at fixed gas fraction, 
 but were almost independent of the merger orbits and orientations.    

In this paper,  we study major and minor merger simulations with primary disc galaxy models of increasing baryonic gas fractions (19\%, 39\%, and 53\%) and decreasing
stellar mass fractions,  but identical in total mass, dark matter concentration, and scale-length.   
The lowest gas-fraction simulations were originally presented in Papers 1 and 2. 
 In \S 2, we describe the simulations, 
the properties of the initial galaxies, and the merger parameters. In \S 3, we briefly
describe the analysis of the resulting simulated images and the criteria for identification as a merger by morphology
and projected separation.  In \S 4, we discuss the resulting observability time-scales and their dependence on the baryonic gas fraction
and merger mass ratios, and we summarize the results and discuss the broader implications in \S 5.    Those familiar with our approach in Papers 1 and 2 
may skip to \S 4$-$5 for the results and discussion.   The simulated $g$ band images and morphology
tables will be available in 2010 at the Multimission Archive at STScI (MAST) as a High-Level Science Products (HLSP) contribution ``Dusty Interacting Galaxy GADGET-SUNRISE
Simulations'' (DIGGSS):  {\bf http://archive.stsci.edu/prepds/diggss} .

\begin{figure*}
\includegraphics[width=168mm]{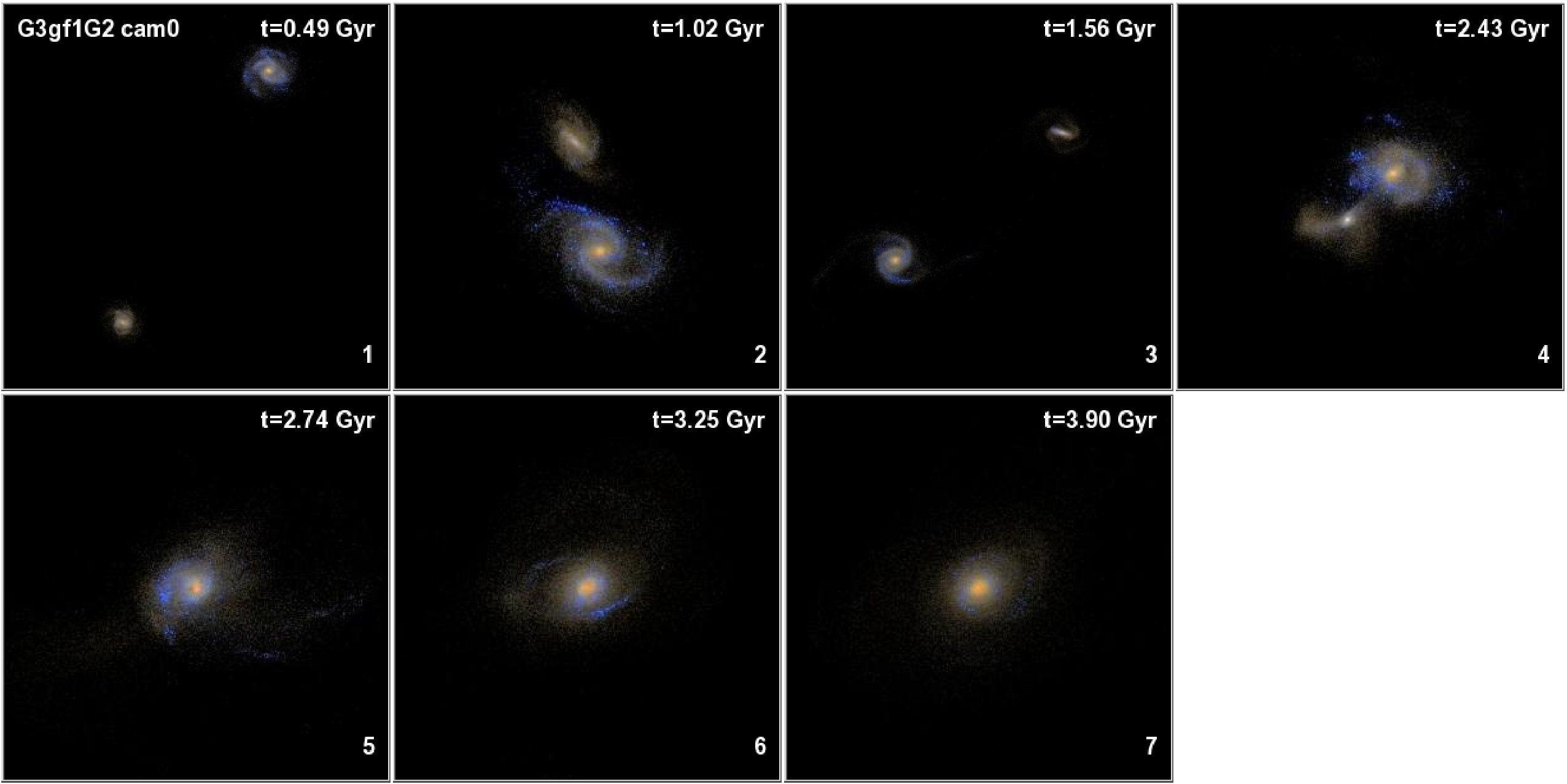}
\caption{ $u-r-z$ composite colour images including dust extinction 
for the G3gf1G2 simulation with $f_{gas}$ = 39\% and 3:1 baryonic mass ratio.  Bright blue star-forming tails and debris
are more evident than in Figure \ref{gf0col} but not Figure \ref{gf1col}.  The viewing angles, merger stages and image
scales are the same as Figure \ref{gf0col}.} \label{gf1col}
\end{figure*}

\begin{figure*}
\includegraphics[width=168mm]{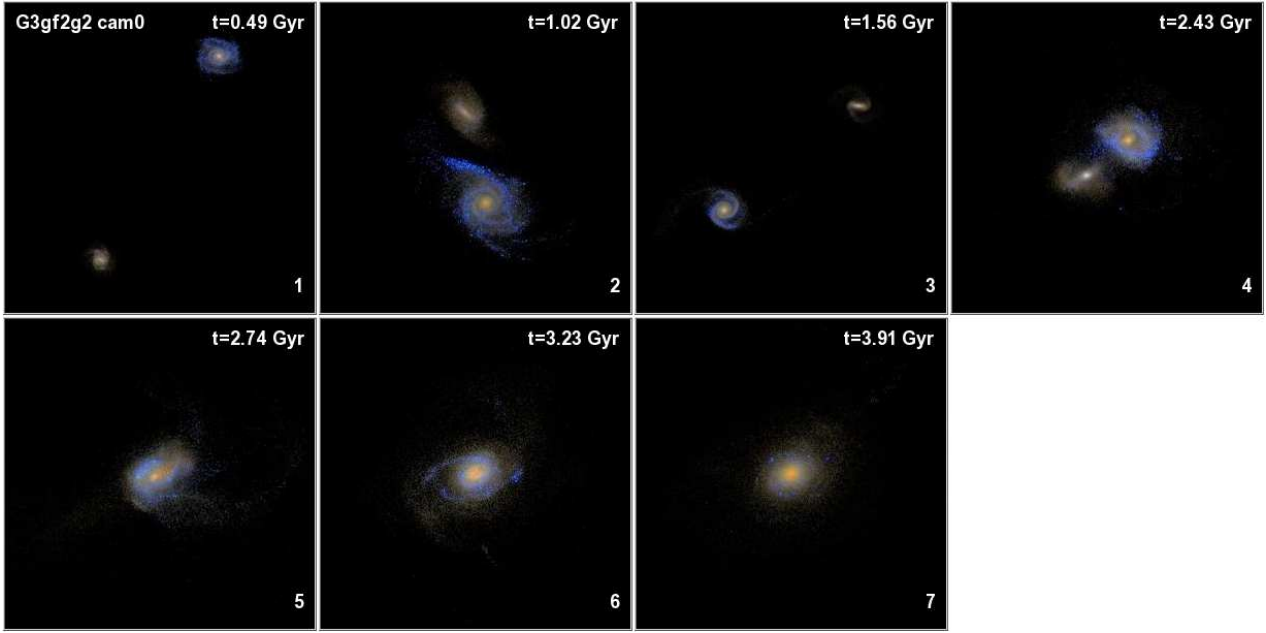}
\caption{  $u-r-z$ composite colour images including dust extinction 
for the G3gf2G2 simulation with $f_{gas}$ = 53\%  and 3:1 baryonic mass ratio.  Bright blue star-forming tails and debris
are more evident than in Figure \ref{gf0col} or \ref{gf1col}. The viewing angles and image
scales are the same as Figure \ref{gf0col} and \ref{gf1col}.} \label{gf2col}
\end{figure*}

\begin{table}
  \centering
  \begin{minipage}{84mm}
    \caption{Merger Mass Ratios}
    \begin{tabular}{@{}ccccc@{}} 
      \hline
      Primary  &  Satellite   & Total   & Stellar  & Baryonic \\
     \hline
      G3       &    G3        & 1:1     & 1:1      & 1:1\\
      G3gf1    &    G3gf1     & 1:1     & 1:1      & 1:1\\
      G3gf2    &    G3gf2     & 1:1     & 1:1      & 1:1\\
      Sbc      &    Sbc       & 1:1     & 1:1      & 1:1\\
     \hline
      G3       &    G2        & 2.3:1   & 3.2:1    & 3.1:1\\
      G3gf1    &    G2        & 2.3:1   & 2.4:1    & 3.1:1\\
      G3gf2    &    G2        & 2.3:1   & 1.9:1    & 3.1:1\\
     \hline
      G3       &    G1        & 5.8:1   & 10.0:1   & 8.9:1\\
      G3gf1    &    G1        & 5.8:1   & 7.6:1    & 8.9:1\\
      G3gf2    &    G1        & 5.8:1   & 5.8:1    & 8.9:1\\
      \hline
\end{tabular}\label{mrtab}
\end{minipage}
\end{table}

\section{Simulations}
Here we briefly describe the galaxy merger simulations and initial conditions.  

\subsection{GADGET N-Body/SPH simulations}
 The details of these simulations, 
their global star-formation histories, and their remnant properties are discussed
in Cox et. al (2004, 2006, 2008). 
All of the simulations presented in this work were performed using the 
N-Body/SPH code {\sc GADGET} (Springel, Yoshida, \& White 2001). 
While we use the first version of GADGET (Springel et al. 2001), the smoothed
particle hydrodynamics are upgraded to use the `conservative entropy' version that
is described in Springel \& Hernquist (2002). Each galaxy is initially modeled as a disc of
stars and gas, a stellar bulge, and a dark matter halo, with the number of particles and masses
for each component given in Table 1.  The stellar and dark matter particles
are collisionless and are subject to only gravitational forces.  The gas particles are also
subject to hydro-dynamical forces. The baryonic and dark matter particles have gravitational
softening lengths of 100 pc and 400 pc respectively.  The SPH smoothing length for the
gas particles indicates the size of the region over which the particle's hydrodynamic quantities are
averaged and is required to be greater than half the gravitational softening length or
$>$ 50 pc.  The radiative cooling rate $\Lambda_{net}$($\rho$, $u$)
is computed for a primordial plasma as described in Katz et al. (1996).

Gas particles are transformed into collisionless star particles assuming the
Kennicutt-Schmidt law (Kennicutt 1998) where the star-formation rate depends
on the local gas density $\rho_{gas}$. 
 These new star particles have typical masses $\sim 10^5$ M$_{\odot}$, and are assigned
ages based on their formation time and metallicities based on the metallicity of
the gas particle from which they are spawned.  We adopt the instantaneous recycling approximation
for metal production whereby massive stars are assumed to instantly become supernovae, 
and the metals produced are put back into the gas phase of the particle. 
In this version of GADGET, metals do not mix and remain in the gas particle in which they
are formed. The enriched gas contribution from stellar winds and Type Ia supernovae are ignored. 
Unlike the metals, there is no recycling of hydrogen and helium to the gas.  

Feedback from supernovae is required to produce stable star-forming discs. 
We adopt a model in which the supernova feedback energy is
dissipated on an 8 Myr time-scale, and have a equation of state  
$P \sim \rho_{gas}^{2}$.  No active galactic nuclei (AGN) are included in these simulations.  As we discussed in Paper 1, 
the exclusion of AGN feedback will not affect the morphological disturbance time-scales calculated here but may
affect the appearance of the merger remnants.

\subsection{SUNRISE Monte Carlo radiative transfer processing}
{\sc SUNRISE} is a parallel code which performs full Monte Carlo radiative-transfer calculations
using an adaptive-mesh refinement grid (Jonsson 2006; Jonsson et al. 2006; Jonsson, Groves, \& Cox 2009).  
We use {\sc SUNRISE v2} to create  $g$-band images at least 30 timesteps for each merger simulation.  
For each {\sc GADGET} simulation timestep,  {\sc SUNRISE} assigns a spectral energy distribution
to each star particle using the STARBURST99 population synthesis models (Leitherer et al. 1999).  

The metallicities of the gas and stars of the initial galaxy models decline
exponentially with the radius of the disc.  The density of dust is linearly proportional
to the density of metals in the gas.  The central metallicities and gradients scale with the
mass of the galaxy, but not the gas fraction.   The details of the initial galaxy metallicities, gradients, and dust extinctions are
given in Rocha et al. (2008). 
 
Given a particular simulation geometry and viewing angle, {\sc SUNRISE v2}
performs the Monte-Carlo radiative transfer calculation for 20 wavelengths from the far-ultraviolet to 
the mid-infrared and interpolates a resulting spectral energy distribution of 510 wavelengths, including the 
effects of absorption and scattering.  Images are created for eleven isotropically positioned viewpoints (``cameras'').
In Figures \ref{gf0col}, \ref{gf1col}, and \ref{gf2col}, 
we show examples of composite  $u-r-z$ images for the G3-G2 merger
simulations with 19\%, 39\%, and 53\% primary galaxy baryonic gas fractions,  viewed face-on (camera 0) at multiple timesteps.

\subsection{Initial Galaxy Models and Merger Parameters}
The galaxy models G3, G2, and G1 have masses, bulge-to-disc ratios, and gas fractions motivated by SDSS estimates of typical local galaxies
 (Table \ref{gparstab}; Cox et al. 2008).  Each galaxy model contains a rotationally supported disc of gas and stars, a non-rotating 
stellar bulge, and a massive dark matter halo (Table \ref{gparstab}). A detailed description of the galaxy disc models can be found in Cox et al. (2006b, 2008), 
Jonsson et al. (2006) and Rocha et al. (2008).  The largest galaxy (G3) is chosen to have a stellar mass $\sim 5 \times 10^{10}$ M$_{\odot}$, 
and the smaller galaxies
are chosen to have stellar masses $\sim 1.5 \times 10^{10}$ M$_{\odot}$ (G2),  and  $5 \times 10^{9}$ M$_{\odot}$ (G1), spanning a factor of 10 in stellar mass.  
The total mass-to-light ratio is assumed to vary with mass such that lower mass galaxy models have higher mass-to-light ratios and the
range in total mass is a factor of 6.   The galaxy merger simulations with the gas fractions tuned to local galaxy values were originally 
presented in Paper 1 (equal-mass merger G3G3) and Paper 2 (unequal-mass mergers G3G2 and G3G1) and are included here for comparison.  

In this work, we explore the effect of gas fraction on the merger morphologies and disturbance time-scales.  Because the satellite G2 and G1 galaxies
contribute relatively little to total gas content and star-formation, we have only modified the baryonic gas fraction of the primary G3 galaxy.  The total
mass,  stellar bulge mass,  overall baryonic fraction, and metallicity have been kept constant.  The central metallicities are $\sim Z_{\odot}$, which may be
an overestimate for high gas-fraction disc galaxies at $z > 1.5$ (Erb et al 2006a, b),  although recent CO observations at $z \ge 1$ suggest
higher gas fractions for high mass, high metallicity discs (Daddi et al. 2009, Tacconi et al. 2009) than derived by Erb et al. (2006a).
 The ratio of disc stars to disc gas has been modified
such that the fraction of total baryons in gas was increased from 19\% for the G3 model to 39\% for the G3gf1 model and 53\% for 
the G3gf2 model. These high gas-fraction discs are merged together in identical equal-mass mergers (G3gf1G3gf1  and G3gf2G3gf2), and with the satellite galaxy
models (G3gf1G2,  G3gf1G1,  G3gf2G2,  G3gf2G1). 

The gas fractions in the G3gf1 and G3gf2 models have been boosted by increasing the gas particle masses and decreasing the disc star particle masses.  
However, as the gas particles are turned into stars,  this results in new star particles with somewhat larger masses than the old star particles.
In order to understand any resulting systematics, we created a modified version of the initial galaxy G3gf2 (G3gf2a) with fewer disc star particles 
and a larger number of gas particles than the G3 model but with gas particles of the same mass as the G3 model.  Any biases should be most
apparent in the highest gas-fraction equal-mass merger because these form the largest number of new star particles.  We compare the equal-mass merger G3gf2aG3gf2a
to the G3gf2G3gf2 merger,  and find no significant difference in the morphologies  (see \S4 for discussion). 
Therefore we conclude that the slightly higher mass new star particles in the G3gf1- and G3gf2- merger simulations do not bias our morphological results. 

The total, baryonic, and stellar mass ratios for all the simulations are given in Table \ref{mrtab}.
Throughout this paper we will generally refer to the baryonic mass ratios of the mergers, but note that the stellar/total mass ratios are
higher/lower than the baryonic mass ratios.   
The G3G3, G3gf1G3gf1, and G3gf2G3gf2 mergers are identical equal-mass mergers with 1:1 mass ratios.  
The G3G2, G3gf1G2, and G3gf2G2 mergers have baryonic mass ratios of $\sim$ 3:1, and hence are more representative of major mergers than rare equal-mass mergers.   
The G3G1, G3gf1G1, and G3gf2G1 mergers have baryonic mass ratios of $\sim$ 9:1 and are considered to be minor mergers.   

All of the simulated G mergers in this paper have the same orbital parameters.   Each pair of galaxies starts on a sub-parabolic orbit with eccentricity = 0.95
and an initial pericentric radius 13.6 kpc.  The galaxies have a roughly prograde-prograde orientation relative to the orbital plane,  with the
primary galaxy tilted 30 degrees from a pure prograde orientation.   In Papers 1 and  2,  we found that the merger orbits and orientations
did not significantly change the time-scales for strong morphological disturbances.  

Finally, we also include the gas-rich prograde-prograde simulation of the Sbc-Sbc merger from Paper 1 (Sbc-Sbc). 
The initial Sbc galaxy model parameters are motivated by observations of local gas-rich, disc-dominated Sbc galaxies similar to 
the Milky Way (see Table 1; Cox et al. 2006b).  The baryonic gas fraction of the Sbc model (52\%) is similar to the G3gf2 model (53\%).  
The Sbc model galaxy has a larger gaseous disc scale-length than the other simulations (16.5 kpc v. 8.6 kpc), therefore the 
majority of its gas lies at large radii.  The Sbc model also has a higher dark matter concentration (11 v. 6 ).  
The Sbc-Sbc merger starts on a parabolic with eccentricity = 1. and an initial pericentric radius of 11 kpc.   The galaxies have a roughly
prograde-prograde orientation,  with galaxy 2 tilted 30 degrees from the plane of the orbit (see Paper 1).  

\begin{figure*}
\includegraphics[width=168mm]{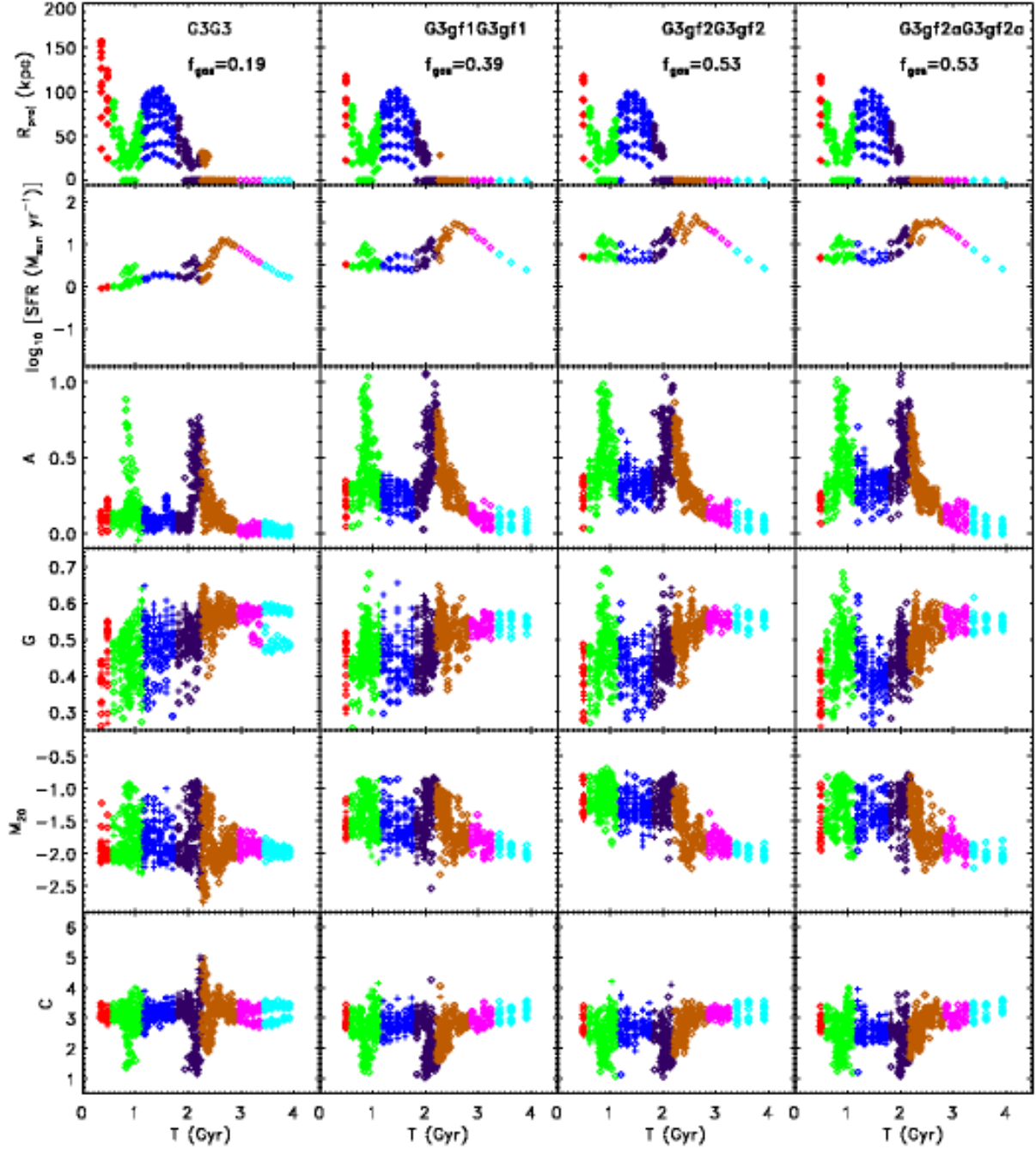}
\caption{Time v. projected separation ($R_{proj}$), log$_{10}$(SFR), and morphology ($A$, $G$, $M_{20}$, $C$) 
for equal-mass mergers G3G3 ($f_{gas}=$ 0.19),  G3gf1G3gf1 ($f_{gas}=$ 0.39),  G3gf2G3gf2 ($f_{gas}=$ 0.53), and G3gf2aG3gf2a ($f_{gas}=$ 0.53).  
The G3gf2G3gf2  and G3gf2aG3gf2a simulations are identical except
for the particle masses/number. The satellite galaxies are shown as crosses, and the primary galaxy/merger is shown as diamonds. The merger stages are colour-coded
with initial galaxies:red, the first pass:green, the maximal separation:blue, second pass(and third pass): purple, final merger:orange, 
post-merger:magenta, and merger remnant at $>$ 1 Gyr after coalescence of the
nuclei as cyan.  }  \label{tgf0}
\end{figure*}

\section{Image Analysis}
We replicate the observations and measurements of real galaxy mergers as closely 
as possible.   We focus on rest-frame $g$ morphologies for purposes of this paper, as
these simulations can be used to calibrate the morphologies of galaxies currently observed from the ground and
with the Hubble Space Telescope in optical and near-infrared wavelengths at $0 < z < 3$. 
In the following section we briefly describe how the simulated $g$ images are degraded and analysed to match real
galaxy morphology measurements; a more detailed discussion may be found in Paper1. 

\subsection{Image degradation}
The $g$ images are produced by {\sc SUNRISE} for each simulation for 11 isotropically positioned viewpoints as a function
of time from $\sim 0.5$ Gyr prior to the first pass to $\ge$ 1 Gyr after the final coalescence in $\sim 30-250$ 
Myr timesteps (depending on the merger state),  up to a maximum runtime of 6 Gyr.  
The field of view of the output images ranges from 200 kpc 
during the initial stages and period of  maximal separation to 100 kpc during the first pass, second pass, 
final merger and post-merger stages.  The intrinsic resolution of the output {\sc SUNRISE} $g$-band images is 333 pc per pixel.

The images output by {\sc SUNRISE} have no background sky noise and no seeing effects, although they
do have particle noise and Monte Carlo Poisson noise. We degrade these images to simulate real data, 
but do not attempt to mimic a particular set of galaxy survey observations.  
We re-bin the images to 105 pc per pixel and convolve the images with a Gaussian function with a FWHM = 400 pc.  
This was done to simulate the effect of seeing but maintain as high spatial resolution as possible.  The values were
chosen to match the Sloan Digital Sky Survey (Abazajian et al. 2003) with 1.5\arcsec\ seeing, 0.396\arcsec\ per pixel plate scale for a galaxy at a distance such
that 1.5\arcsec\ $\sim$ 400 pc.  We also add random Poisson noise background to simulate sky noise but scale this noise
to maintain a high signal-to-noise for the primary galaxies ($>20$ per pixel within the Petrosian radius).  

\begin{figure}
\includegraphics[width=84mm]{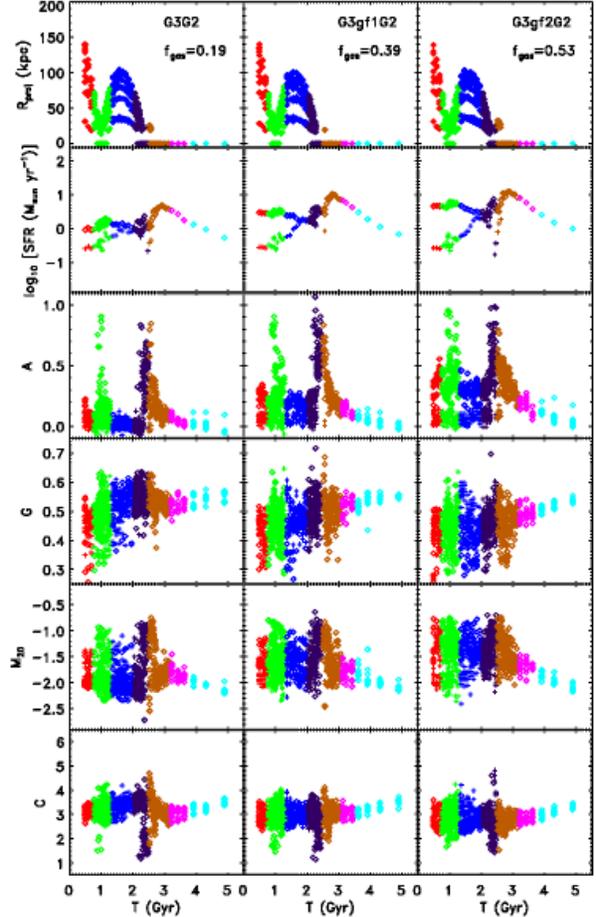}
\caption{Time v. projected separation ($R_{proj}$), log$_{10}$(SFR), and morphology ($A$, $G$, $M_{20}$, $C$) 
for 3:1 baryonic mass ratio mergers G3G2,  G3gf1G2,  and G3gf2G2.  The satellite galaxies are shown as crosses, and the primary galaxy/merger is shown as diamonds.
The merger stages are colour-coded as in the previous plot. }  \label{tgf1}
\end{figure}

\begin{figure}
\includegraphics[width=84mm]{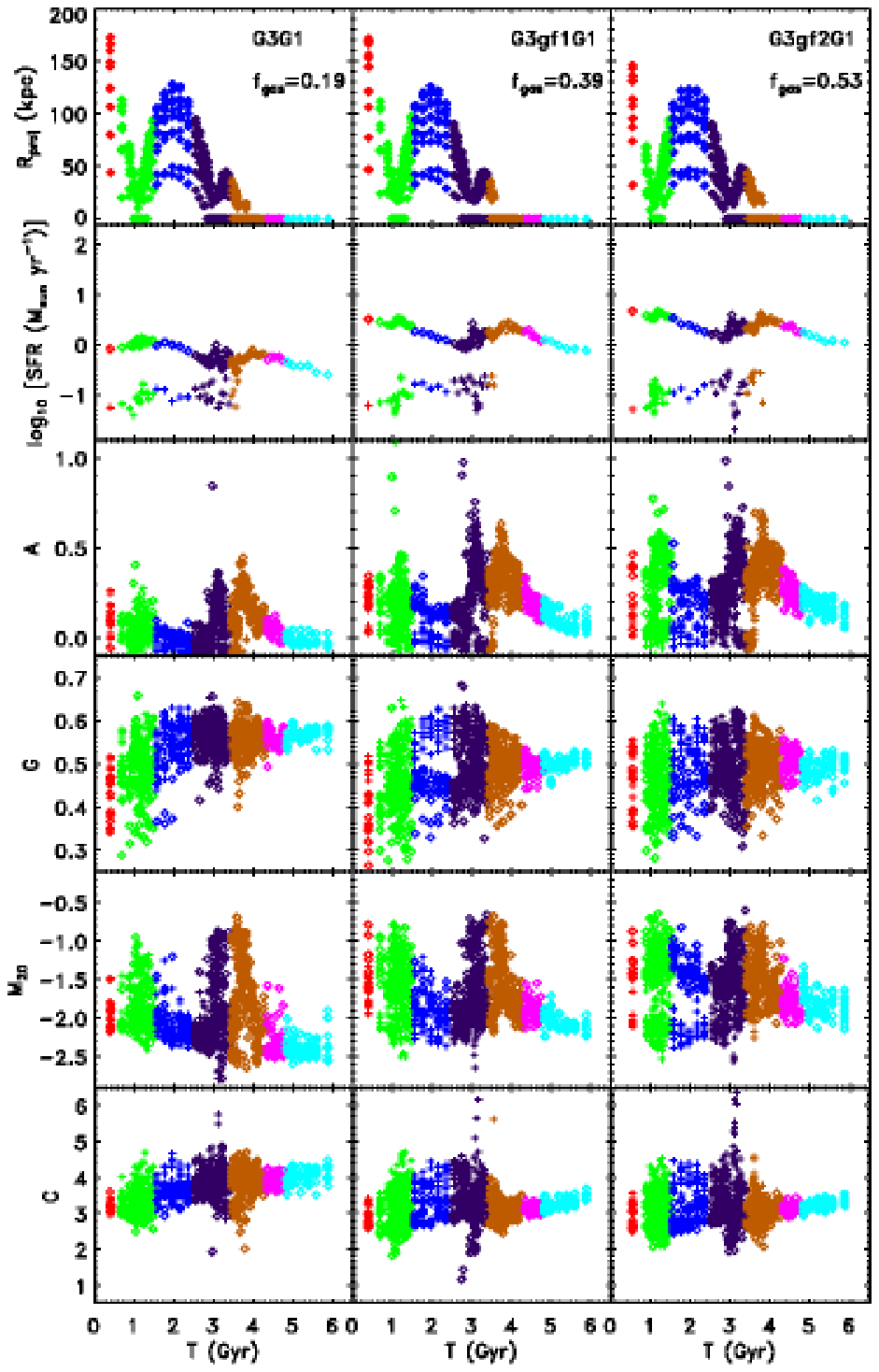}
\caption{Time v. projected separation ($R_{proj}$), log$_{10}$(SFR), and morphology ($A$, $G$, $M_{20}$, $C$) 
for 9:1 baryonic mass ratio mergers G3G1,  G3gf1G1,  and G3gf2G1. The satellite galaxies are shown as crosses, and the primary galaxy/merger is shown as diamonds.
 The merger stages are colour-coded as in the previous plots. }  \label{tgf2}
\end{figure}
\subsection{Morphology Measurements}
Each image is run through an automated galaxy detection algorithm integrated into our 
IDL code. If the centres of the merging galaxies are less than 10 kpc apart, they are generally detected as a single object. 
If 2 distinct galaxies are detected, the detection segmentation maps are used to mask out the other galaxy while each 
galaxy’s morphology is measured.   The projected separation $R_{proj}$ is measured when two galaxies are detected.
For each detected object, we calculate the Petrosian radii in circular and elliptical apertures, concentration $C$, 180 degree rotational 
asymmetry $A$, 
the Gini coefficient $G$, and the second-order moment of the brightest 20\% of the light $M_{20}$. (Please refer to 
Lotz, Primack, \& Madau 2004, Conselice 2003, and Paper 1 for detailed definitions).  

In Paper 1, we studied the effect of numerical resolution on the simulation morphologies.  We found small but significant differences in 
the average $M_{20}$ and $A$ values when we compared the standard numerical resolution simulation to simulations with 4 and 10 times as many
particles, and corrected for these offsets. Because the numerical resolution of the simulations presented here are similar to the 
standard resolution simulations in Paper 1  ($\sim 10^5$ particles), we apply the same correction of $\delta M_{20} = -0.157$ and $\delta A = -0.115$ to 
the values in this work.

\begin{figure*}
\includegraphics[width=168mm]{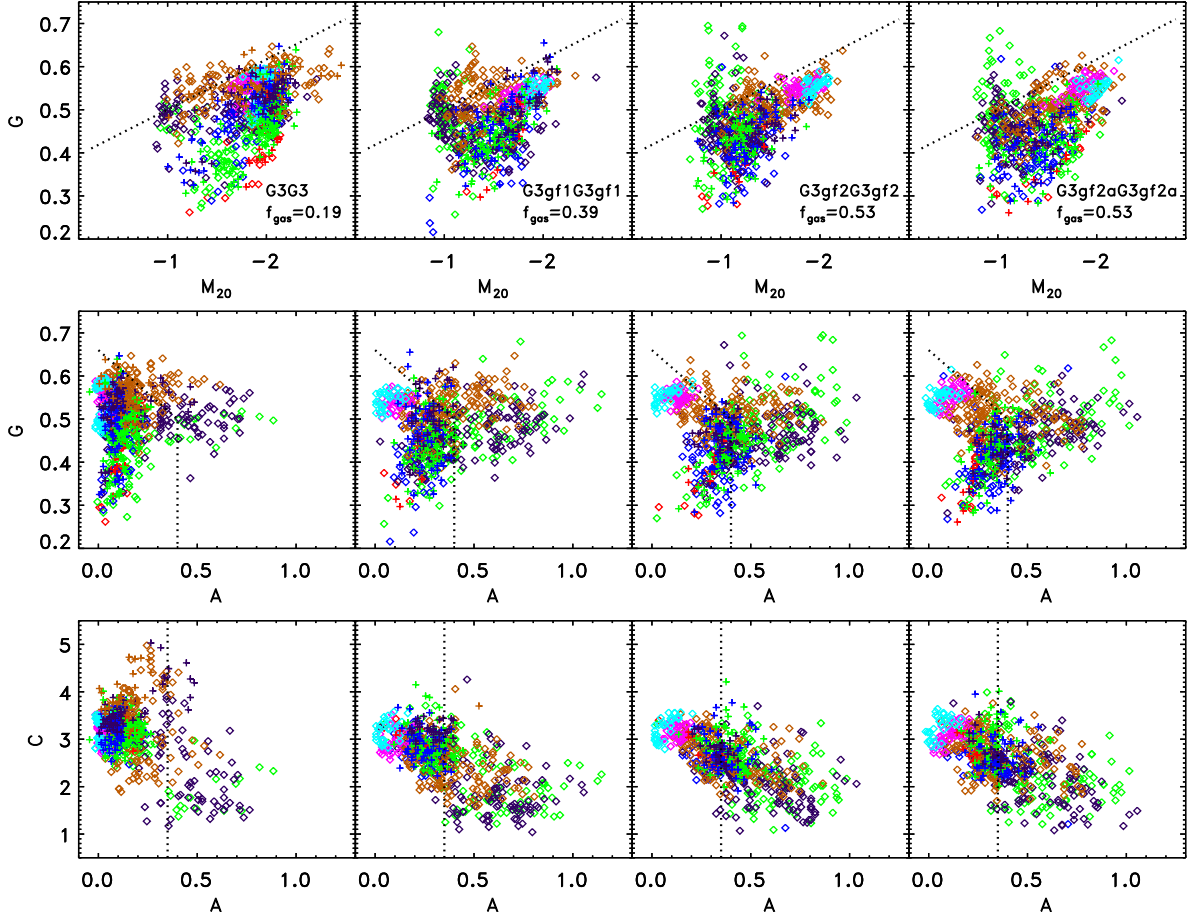}
\caption{$G-M_{20}$, $G-A$, and $C-A$ for  the equal-mass mergers G3G3 ($f_{gas}=$ 0.19),  G3gf1G3gf1 ($f_{gas}=$ 0.39),  
G3gf2G3gf2 ($f_{gas}=$ 0.53), and G3gf2aG3gf2a ($f_{gas}=$ 0.53).  
The G3gf2G3gf2  and G3gf2aG3gf2a simulations are identical except
for the particle masses/numbers. The satellite galaxies are shown as crosses, and the primary galaxy/merger is shown as diamonds.
Objects are classified as morphologically-disturbed merger candidates if they fall above ($G-M_{20}$, top panels)
or to the right ($G-A$, $A$, lower panels) of the dashed lines.
The merger stages are colour-coded as in the previous plots.  } \label{morphgf0}
\end{figure*}

\begin{figure}
\includegraphics[width=84mm]{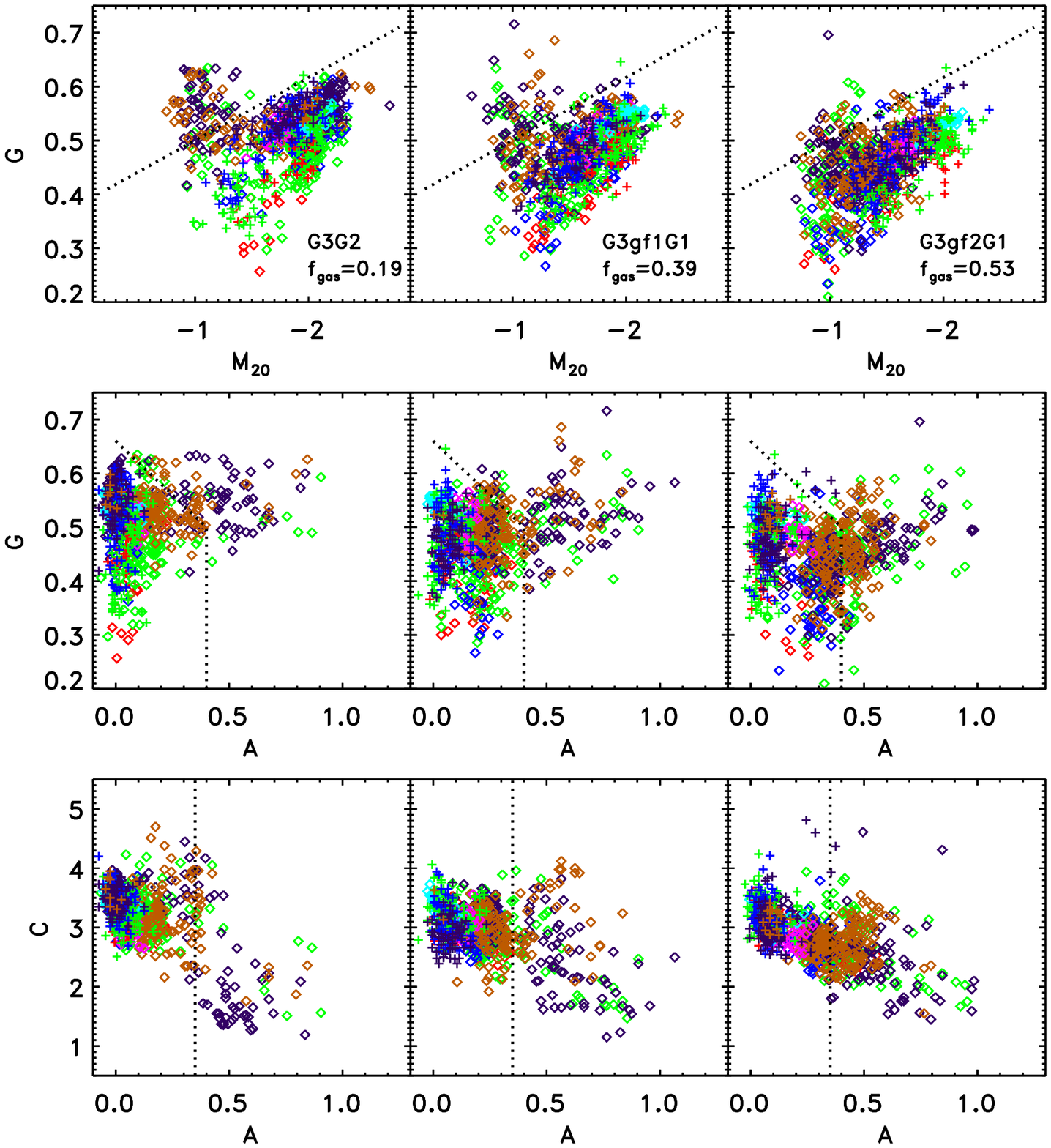}
\caption{$G-M_{20}$, $G-A$, and $C-A$ for 3:1 baryonic mass ratio mergers G3G2,  G3gf1G2,  and G3gf2G2.
Figure symbols and lines are the same as Figure \ref{morphgf0}. } \label{morphgf1}
\end{figure}

\subsection{Merger Classification and Time-Scales}
Lotz et al. (2004) found that local ultra-luminous galaxies visually classified as mergers could be distinguished from
the sequence of normal Hubble type galaxies with  
\begin{equation}
G > -0.115\ M_{20} + 0.384
\end{equation}
or
\begin{equation}
G  > -0.4\ A + 0.66\ {\rm or}\  A \ge 0.4
\end{equation}

Asymmetry alone is also often used to classify merger candidates.   The calibration of local mergers by
Conselice (2003) finds the following merger criterion:
\begin{equation}
A \ge 0.35
\end{equation} 

Galaxies at higher redshift cannot be imaged at as high spatial resolution as local galaxies even
when observed with $HST$.  The measured morphologies of galaxies at $z > 0.25$ imaged with $HST$ will have
non-negligible biases as a result of this lower spatial resolution (Lotz et al. 04).  See Paper 1 for a
discussion of how the morphologies and merger criteria change at $z \sim 1$ $HST$ resolution, and
how the time-scale computed here may be applied to $HST$ data.  

Close kinematic pairs are also probable pre-merger systems.  
We assume $h=0.7$ and we estimate the time-scales during which merging galaxies can be found as separate objects
within $5  < R_{proj} < $ 20, 10 $ < R_{proj} <$ 30, 10 $ < R_{proj} < $ 50, 
and 10 $< R_{proj} < $ 100 $h^{-1}$ kpc.    The simulated merging galaxies always have relative velocities $<$ 500  
km s$^{-1}$. 

We calculate each simulation's average observability time-scale for the $G-M_{20}$, $G-A$, and $A$ criteria given
above by averaging the results of the 11 isotropic viewpoints.  
Because we wish to determine the number density of merger events rather than the number of galaxies
undergoing a merger, galaxies that have not yet merged but identified morphologically as merger candidates are 
weighted accordingly.  The time that each pre-merger galaxy is morphologically disturbed is summed (not averaged)
to the time that the post-merger system appears disturbed. 
No such weighting is done for the close pair time-scales as this factor is generally
included in the merger rate calculation (e.g. Patton et al. 2000).

\begin{figure}
\includegraphics[width=84mm]{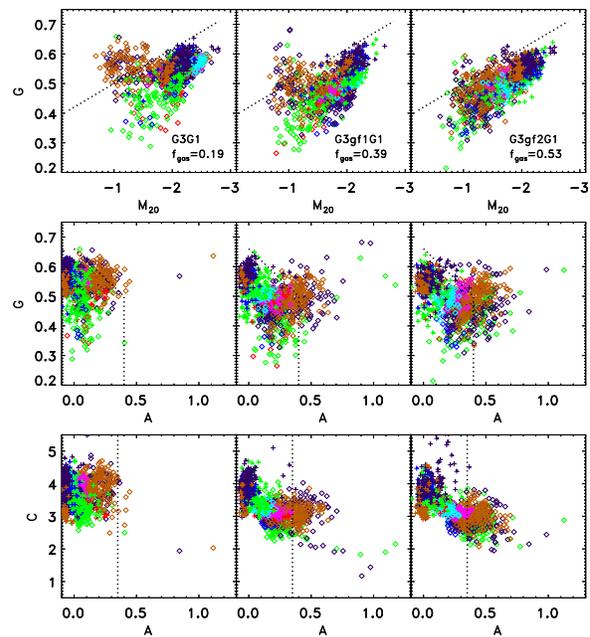}
\caption{$G-M_{20}$, $G-A$, and $C-A$ for 9:1 baryonic mass ratio  mergers G3G1,  G3gf1G1,  and G3gf2G1.
Figure symbols and lines are the same as Figure \ref{morphgf0}.  } \label{morphgf2}
\end{figure}

\begin{figure}
\includegraphics[width=84mm]{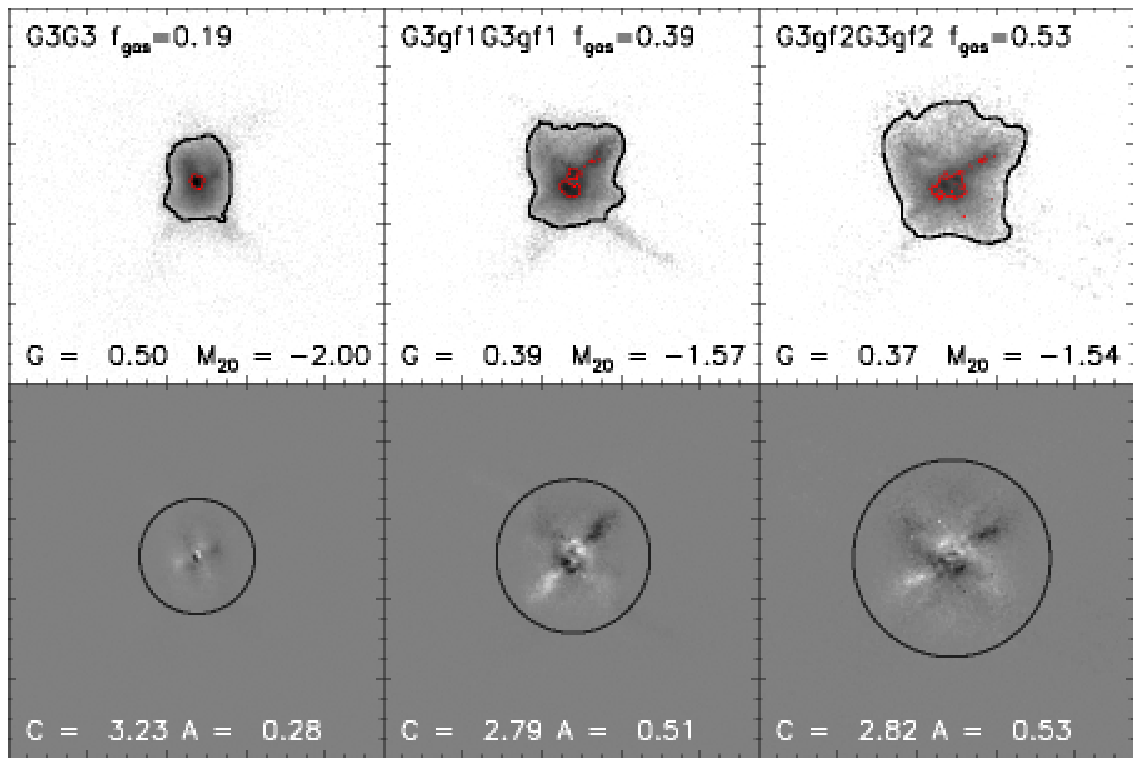}
\caption{Equal-mass mergers G3G3, G3gf1G3gf1, and G3gf2G3gf2 viewed by camera 4 just after the coalescence of the nuclei at t=2.5 Gyr.
None of the simulated images are classified as mergers by $G-M_{20}$ because they have a single nucleus, 
but the images of G3gf1G3gf1 and G3gf2G3gf2 are classified as mergers by $A$ and $G-A$ because of the increased number and brightness of the 
star-forming regions.  
{\it Top:} The pixels containing the brightest 20\% of the flux (used to calculate $M_{20}$) are shown by the inner red contours; the pixels used to 
measure $G$ are marked by the
outer black contours.  {\it Bottom: } The difference image between the original and 180 degree rotated image is shown.  The circles show the 
1.5 x Petrosian radius within which $A$ is measured. } \label{stamps}
\end{figure}

\section{Results}
\subsection{Evolution with Merger Stage}
  The time evolution of each simulation depends on the mass ratio of the galaxies, therefore we
compare the behavior of each simulation as a function of merger stage rather than time. 
We define seven different merger stages based on the positions of the galaxy nuclei in real space,  
which are colour-coded in Figures \ref{tgf0} $-$ \ref{morphgf2}. These are the initial encounter (red), 
the first pass (green),  maximal separation immediately after the first pass (blue), the second pass or final approach (purple),
the final merger (orange), the post-merger observed 0.5$-$1 Gyr after the final merger (magenta), and the remnant
observed $>$ 1 Gyr after the final merger (cyan). 

In Figures \ref{tgf0}, \ref{tgf1}, and \ref{tgf2} we examine the projected separations $R_{proj}$, quantitative morphologies 
($G$, $M_{20}$, $C$, and $A$), and star-formation rate per object as a function of time and merger stage.  Each figure shows a 
fixed merger mass ratio and increasing  baryonic gas fraction for the primary galaxy. 
 The scatter in $R_{proj}$ and morphologies at each timestep is the variation
in the merger appearance with the 11 viewing angles. The minor mergers (G3G1, G3gf1G1, G3gf2G1) take significantly longer 
to complete than the major mergers, and experience a third and final pass before the galaxies coalesce.  

The simulations show similar patterns of starbursts and morphological disturbances with merger stage.  
The initial segmentation maps computed to identify each galaxy are used to compute the total star-formation rate for each galaxy
at each timestep and camera. The typical star-formation rate of the initial primary discs and mergers increases with $f_{gas}$ of the 
primary galaxies (Figures \ref{tgf0} $-$ \ref{tgf2}).  
The major mergers experience a burst of star-formation at the first pass and a few hundred Myr after the strongest morphological disturbances as the nuclei merge 
 (see also Papers 1 and 2).  The minor mergers also have enhanced star-formation rates at the close passages,  
although they do not show strong starbursts relative to the initial star-formation rate at the final merger (Figure \ref{tgf2}; see also Cox et al. 2008).  

The baryonic gas fraction $f_{gas}$ of the primary galaxy strongly affects the morphology of the merging system.      
All of the mergers show peaks in $A$ and $M_{20}$ and minima in $C$ during the close passages and just before the final merger.
The higher gas-fraction simulations (G3gf1, G3gf2) have higher $A$ at the close passages and final merger than the lowest gas-fraction
simulations.    These simulations also have higher $A$ between the first pass and final merger stages,  often with values 
greater the merger criterion ($A \ge 0.35$). The $M_{20}$ values for the extremely gas-rich primary galaxies are significantly higher in the
early stages of the merger.  $G$ is similar for the initial discs,  but is suppressed to lower values between the 
first pass and final merger for the extremely gas-rich primary galaxies. $C$ is relatively unaffected by
gas fraction,  although the high $f_{gas}$ simulations are less likely to achieve high $C$ values.   

The higher gas fraction simulations have higher star formation rates and dust production. The higher star formation rates result in more bright star-forming knots 
in the outer regions, which contributes to higher $M_{20}$ and $A$ and lower $C$.  Higher star formation rates also result in higher gas metallicities, whose
primary effect is to increase the amount of dust.   We found in Paper 1 that both $G$ and $M_{20}$ individually are more
sensitive to dust extinction than $A$ and $C$. Dust preferentially extinguishes the central nuclei,  which lowers $G$ and $C$ and raises $M_{20}$.      

The morphologies of the major merger remnants (cyan points, Figure \ref{tgf0} $-$ \ref{tgf1}) are not 
strongly affected by the initial gas fractions, and are consistent with bulge-dominated disc galaxies (e.g. Sb).  The minor merger remnants
are more likely to remain disc-like if they have higher gas fractions (Figure \ref{tgf2}).   However, as we discuss in Paper 2, the minor merger simulations
take considerably longer to coalesce and consume more of their gas during the course of the simulations.   We concluded that the spheroid-like $G$, $M_{20}$, and $C$
values for the lowest gas-fraction minor merger (G3G1) were the result of the long simulation runtime (6 Gyr) rather than the dynamics of the merger remnant. 
We will study the remnant properties in more detail in a future paper.    

As we mentioned in \S2,  the gas fractions of the G3gf1 and G3gf2 model were increased by increasing the gas particle masses and decreasing the
disc star particle mass relative to the G3 model.  Therefore new star particles formed in the G3gf1 and G3gf2 simulations have slightly
larger masses than the G3 simulations.  In Paper 1,  we found that low numerical resolution -- i.e. high particle masses -- could systematically 
bias the $A$ and $M_{20}$ values.    To determine if the high gas fraction model suffered the same problem, 
 we simulated a modified version of the G3gf2a model  where the gas fraction was increased by changing the number of gas and disc star particles.    
 We compare the morphologies of an equal-mass
G3gf2aG3gf2a merger to the G3gf2G3gf2 merger in Figures \ref{tgf0} and \ref{morphgf0}.  These simulations have the highest star-formation rates, and therefore 
produce the largest number of new star particles.  We find no difference between the two simulations,  and conclude that the 
slightly larger new particle masses of the high gas fraction simulations do not effect our results.

\begin{table}
  \centering
  \begin{minipage}{84mm}
    \caption{Morphological Disturbance Time-Scales}
    \begin{tabular}{@{}lccc@{}}
      \hline
      Simulation &  T($G-M_{20}$)  & T($G-A$)  & T($A$) \\
      &   (Gyr)                &   (Gyr)          & (Gyr)  \\ 
      \hline
      \multicolumn{4}{c}{Equal-mass mergers} \\
      \hline    
      SbcSbc      &  0.39$\pm$ 0.16 &  0.78$\pm$ 0.21 &  0.74$\pm$ 0.17   \\
      G3G3       &  0.16$\pm$ 0.07 &  0.31$\pm$ 0.08 &  0.23$\pm$ 0.11   \\
      G3gf1G3gf1 &  0.24$\pm$ 0.14 &  0.81$\pm$ 0.14 &  0.76$\pm$ 0.16   \\
      G3gf2G3gf2 &  0.24$\pm$ 0.16 &  1.25$\pm$ 0.20 &  1.54$\pm$ 0.28   \\
      G3gf2aG3gf2a &  0.16$\pm$ 0.06 &  1.00$\pm$ 0.24 &  1.32$\pm$ 0.32   \\
      \hline
      \multicolumn{4}{c}{3:1 baryonic mass ratio mergers} \\
      \hline   
      G3G2    &  0.25$\pm$ 0.08 &  0.30$\pm$ 0.13 &  0.24$\pm$ 0.11   \\
      G3gf1G2 &  0.26$\pm$ 0.09 &  0.50$\pm$ 0.08 &  0.47$\pm$ 0.08   \\
      G3gf2G2 &  0.14$\pm$ 0.09 &  0.80$\pm$ 0.11 &  1.04$\pm$ 0.20   \\
      \hline
      \multicolumn{4}{c}{9:1 baryonic mass ratio mergers} \\
      \hline   
      G3G1    &  0.36$\pm$ 0.15 &  0.27$\pm$ 0.13 &  0.03$\pm$ 0.03   \\
      G3gf1G1 &  0.21$\pm$ 0.08 &  0.60$\pm$ 0.26 &  0.66$\pm$ 0.19   \\
      G3gf2G1 &  0.14$\pm$ 0.09 &  1.15$\pm$ 0.44 &  1.30$\pm$ 0.34   \\
      \hline
    \end{tabular}\label{tmorphtab}
  \end{minipage}
\end{table}

\subsection{Merger Diagnostics}
In Figures \ref{morphgf0} $-$ \ref{morphgf2},  we plot the simulation morphologies on the merger diagnostic plots $G-M_{20}$, $G-A$, and $C-A$.   The lowest 
$f_{gas}$ major mergers (G3G3, G3G2) are identified just before the final merger (purple points) with $A$ and at the final merger (orange points) in $G-M_{20}$ 
(Figures \ref{morphgf0} $-$ \ref{morphgf1}).  
The higher $f_{gas}$ major mergers are also identified at the first pass (green points) by all three methods,  and are somewhat less likely to be
identified at the final merger via $G-M_{20}$.   The lowest $f_{gas}$ minor merger G3G1 is identified  by $G-M_{20}$ and $G-A$ at the second pass and
the final merger (purple and orange points),  but does not meet the $A > 0.35$ merger criteria (Figure \ref{morphgf2}).  
As $f_{gas}$ is increased,  minor mergers are more likely to be detected by $A$ and $G-A$ around the final merger but slightly less 
likely to be found via $G-M_{20}$.   

To illustrate the different behaviors of $G-M_{20}$ and $C-A$,  we show the equal-mass mergers just after the final merger 
at t=2.5 Gyr in Figure \ref{stamps}.  After the final merger,  the object has a single central nucleus and
spatially extended regions of star-formation which appear brighter and larger as $f_{gas}$ increases. 
None of the simulated images are classified as mergers by $G-M_{20}$ because they have a single nucleus, 
but the images of the G3gf1G3gf1 and G3gf2G3gf2 simulations are classified as mergers by $A$ and $G-A$ because of the increased number 
and brightnesses of the star-forming regions.  As $f_{gas}$ increases, the Petrosian
radius and the regions used to measure $G$, $M_{20}$, $C$, and $A$ become larger.  (These measures are designed to be
independent of size, therefore are {\it not} affected by physical size alone -- see Lotz et al. 2004).   
The lower panels show the difference image between the original image and the image rotated by 180 degrees.  Asymmetry is measured using the pixels
within the circular apertures with radii equal to 1.5 $\times$ the Petrosian radius (circles, lower panels).    As $f_{gas}$ increases, the residuals and
asymmetries increase to exceed the merger criterion.

\begin{figure*}
\includegraphics[width=168mm]{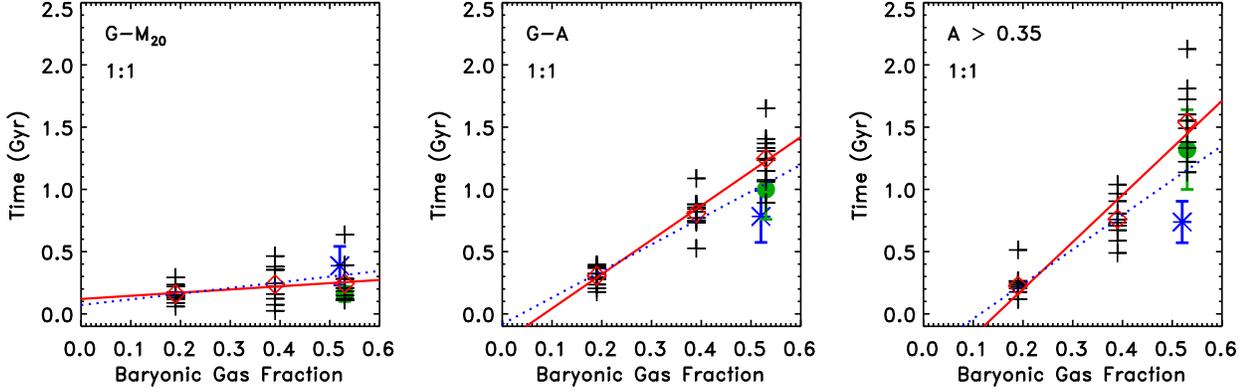}
\caption{Time-scales for morphological disturbances in $G-M_{20}$, $G-A$, and $A$ v. baryonic gas fraction for the equal-mass mergers. 
The black crosses show the time-scales for each viewing angle and the red diamonds are the average time-scales.  The average time-scales 
for lower gas particle mass simulation G3gf2aG3gf2a (green circles) are consistent with the G3gf2G3gf2 time-scales.  
The average $G-A$ and $A$ time-scales for a equal-mass disc-disc merger with larger gas disc scale-length
and f$_{gas}$ = 52\% (SbcPP; blue stars)  are lower than the G3gf2G3gf2 simulation time-scales. The linear fits for $T$ v. $f_{gas}$ are shown for
the 1:1 G mergers only (solid red lines) and 1:1 G and Sbc mergers (dashed blue lines; Table \ref{fittab}). } \label{tmorphgf0}
\end{figure*}

\begin{figure*}
\includegraphics[width=168mm]{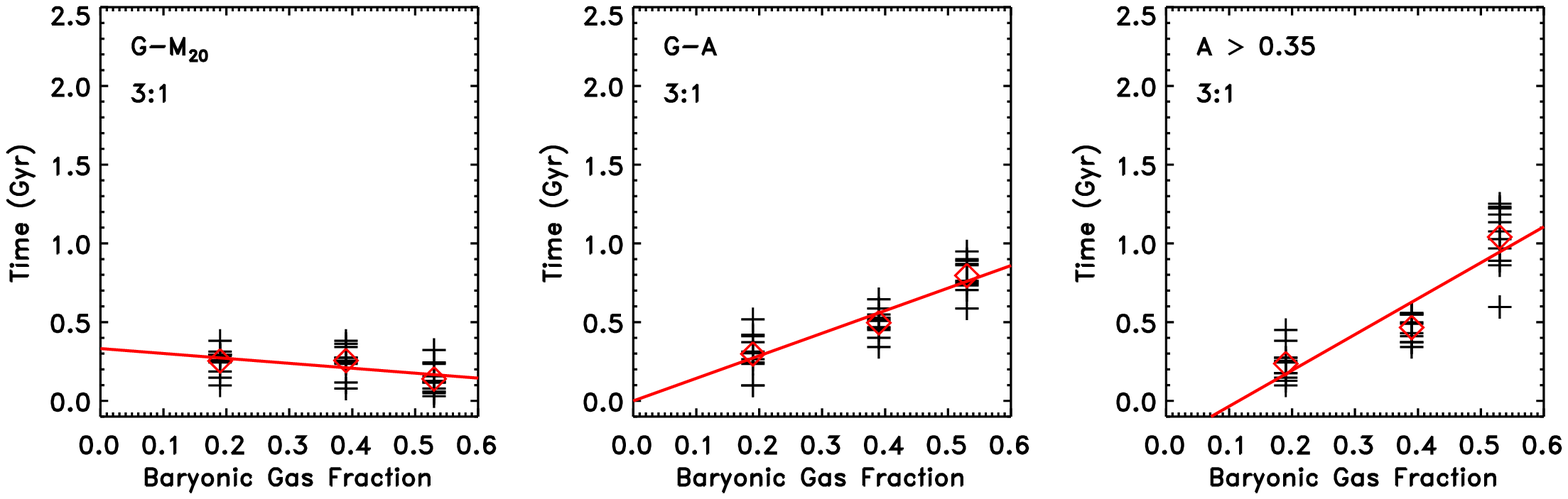}
\caption{Time-scales for morphological disturbances in $G-M_{20}$, $G-A$, and $A$ v. baryonic gas fraction for the 3:1 baryonic mass ratio mergers. 
The black crosses show the time-scales for each viewing angle  and the red diamonds are the average time-scales The linear fits for $T$ v. $f_{gas}$ are shown for
the 3:1 mergers (solid red lines; Table \ref{fittab}).} \label{tmorphgf1}
\end{figure*}

\begin{figure*}
\includegraphics[width=168mm]{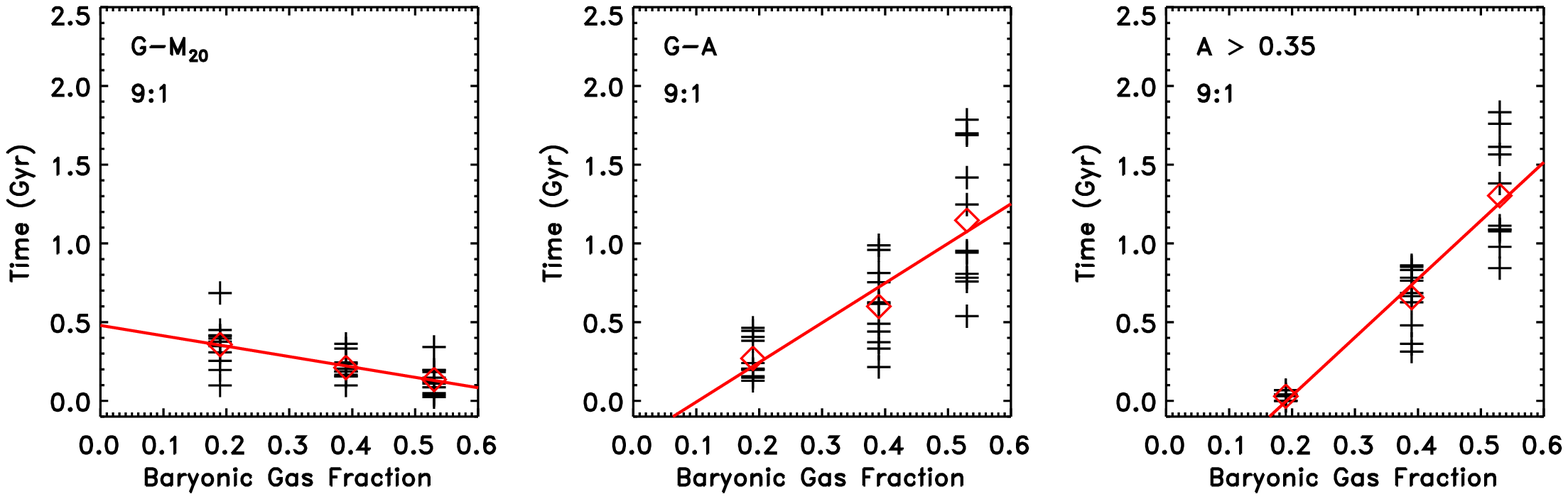}
\caption{Time-scales for morphological disturbances in $G-M_{20}$, $G-A$, and $A$ v. baryonic gas fraction for the 9:1 baryonic mass ratio mergers. 
The black crosses show the time-scales for each viewing angle  and the red diamonds are the average time-scales. The linear fits for $T$ v. $f_{gas}$ are shown for
the 9:1 mergers (solid red lines; Table \ref{fittab}}.) \label{tmorphgf2}
\end{figure*}

\begin{table}
  \centering
  \begin{minipage}{84mm}
    \caption{Observability Time-scales v. $f_{gas}$ }
  \begin{tabular}{@{}lcc@{}}
\hline
Morphology & $T_0$\footnote{T = $T_0$ + Slope $\times$ $f_{gas}$} & Slope$^a$ \\
           & (Gyr) & (Gyr per unit $f_{gas}$) \\
\hline
\multicolumn{3}{c}{Equal-mass mergers} \\
\hline 
$G-M_{20}$    &   0.12 $\pm$ 0.06    &   0.25 $\pm$ 0.16 \\
 w/ Sbc \footnote{Fit includes Sbc-Sbc simulation}      &   0.07 $\pm$ 0.07  &   0.46 $\pm$ 0.16\\
$G-A$        &  -0.23 $\pm$ 0.07   &  2.75 $\pm$ 0.18\\
 w/ Sbc$^b$      &  -0.08 $\pm$ 0.11  &   2.13 $\pm$ 0.25\\
$A$          &  -0.56 $\pm$ 0.11    &  3.78 $\pm$ 0.28\\
 w/ Sbc$^b$      &    -0.32 $\pm$ 0.11  &   2.78 $\pm$ 0.25\\
\hline
\multicolumn{3}{c}{3:1 baryonic mass ratio mergers} \\
\hline 
$G-M_{20}$  &   0.33 $\pm$ 0.05    &  -0.31 $\pm$ 0.12\\
$G-A$   &   0.00 $\pm$ 0.06    &  1.43 $\pm$ 0.15\\
$A$  &  -0.26 $\pm$ 0.09    &  2.28 $\pm$ 0.23\\
\hline
\multicolumn{3}{c}{9:1 baryonic mass ratio mergers} \\
\hline 
$G-M_{20}$  &   0.48 $\pm$ 0.05    &  -0.66 $\pm$ 0.14 \\
$G-A$   &  -0.26 $\pm$ 0.15    &   2.52 $\pm$ 0.39\\
$A$  &  -0.71 $\pm$ 0.11    &  3.71 $\pm$ 0.29\\
\hline
\end{tabular}\label{fittab}
\end{minipage}
\end{table}

\begin{table*}
  \centering
  \begin{minipage}{168mm}
  \caption{Close Pair Time-Scales}
  \begin{tabular}{@{}lcccc@{}}
\hline
Simulation &  T($5<R_{proj} < 20$\footnote{$R_{proj}$ has units kpc $h^{-1}$.})  & T($10 < R_{proj} < 30^a$)  & T($10 < R_{proj} < 50^a$)  & T ($10 < R_{proj} < 100^a$) \\
&   (Gyr)                &   (Gyr)          & (Gyr)  & (Gyr)  \\ 
\hline
\multicolumn{5}{c}{Equal-mass mergers} \\
\hline 
SbcSbc      &    0.15$\pm$   0.19 &    0.35 $\pm$ 0.23  &    0.90 $\pm$ 0.16  &    1.20 $\pm$ 0.18   \\  
G3G3       &    0.39$\pm$   0.30 &    0.72$\pm$   0.39 &    1.21$\pm$   0.38 &    1.85$\pm$   0.13 \\
G3gf1G3gf1 &    0.20$\pm$   0.38 &    0.47$\pm$   0.36 &    0.94$\pm$   0.34 &    1.58$\pm$   0.15 \\
G3gf2G3gf2 &    0.13$\pm$   0.26 &    0.43$\pm$   0.29 &    0.89$\pm$   0.32 &    1.51$\pm$   0.17 \\
G3gf2aG3gf2a & 0.14$\pm$   0.27 &    0.40$\pm$   0.28 &    0.88$\pm$   0.34 &    1.50$\pm$   0.21  \\
\hline
\multicolumn{5}{c}{3:1 baryonic mass ratio mergers} \\
\hline 
G3G2    &    0.26$\pm$   0.15 &    0.59$\pm$   0.44 &    1.15$\pm$   0.36 &    1.98$\pm$   0.16 \\
G3gf1G2 &    0.25$\pm$   0.21 &    0.60$\pm$   0.47 &    1.16$\pm$   0.40 &    1.95$\pm$   0.12 \\
G3gf2G2 &    0.35$\pm$   0.23 &    0.72$\pm$   0.45 &    1.27$\pm$   0.34 &    2.07$\pm$   0.14 \\
\hline
\multicolumn{5}{c}{9:1 baryonic mass ratio mergers} \\
\hline 
G3G1    &    0.44$\pm$   0.28 &    0.96$\pm$   0.47 &    1.67$\pm$   0.62 &    2.87$\pm$   0.31 \\
G3gf1G1 &    0.41$\pm$   0.21 &    0.95$\pm$   0.49 &    1.64$\pm$   0.55 &    2.83$\pm$   0.31 \\
G3gf2G1 &    0.46$\pm$   0.24 &    1.00$\pm$   0.53 &    1.74$\pm$   0.53 &    2.98$\pm$   0.33 \\
\hline
\end{tabular}\label{pairtab}
\end{minipage}
\end{table*} 

\subsection{Observability Time-scales v. $f_{gas}$}
We calculate the time each simulation is classified as a merger using morphological criteria given in Equations 9$-$11.
These time-scales (black crosses) are plotted in Figures \ref{tmorphgf0}$-$\ref{tmorphgf2},  and the viewing-angle averaged values and standard
deviations are given in Table 3.  
We fit the dependence of the observability time-scale on $f_{gas}$,  and give these fits in Table \ref{fittab}.   Again, we find no
significant difference between the G3gf2G3gf2  and G3gf2aG3gf2a simulations (green circles; Table \ref{tmorphtab}). 

The time-scale for detecting a merger with high asymmetry is a strong function of the baryonic gas fraction of the primary galaxy, 
regardless of the merger mass ratio. For the major mergers,  the $A$ and $G-A$ time-scales roughly doubles from $\sim$ 200$-$300 Myr 
to $\sim$ 500$-$700 Myr when $f_{gas}$ is increased 
from 19\% to 39\%,  and roughly doubles again to $\sim$ 1$-$1.4 Gyr when $f_{gas}$ is increased from 39\% to 53\%.   

The asymmetry behavior of the minor mergers is even more dramatic because the lowest $f_{gas}$ minor merger, G3G1,  does not show $A > 0.35$. 
The  minor merger observability time-scales increase from $\sim$ 0 at $f_{gas}$ = 19\% to 1.3 Gyr at $f_{gas} =$ 53\%.   
The minor merger G3gf2G1 simulation $A$ and $G-A$ time-scales are as long as the equal-mass G3gf2G3gf2 simulation because it experiences
an additional peak in asymmetry at the second pass as well as at the first pass and  the final merger.  Given these trends, it is possible 
that even higher mass ratio ($>$ 9:1) high gas-fraction ($f_{gas} >$ 20\%) mergers may be detected by asymmetry.

The $G-M_{20}$ observability time-scale,  on the other hand,  does not appear to be a strong function of gas fraction nor mass ratio. 
There is a slight trend towards {\it shorter} time-scales with higher gas fraction for unequal-mass mergers because the
merging nuclei are more likely to be obscured by dust in the high gas-fraction simulations.  In general, the $G-M_{20}$ time-scales
have a much weaker correlation with $f_{gas}$ than $A$,  as the increased $M_{20}$ values are countered by the decreased $G$ values. 
In Paper 2, we found that a 39:1 baryonic mass ratio merger with local gas fraction was not classified as merger by $G-M_{20}$.
Given the inverse correlation of T($G-M_{20}$) with $f_{gas}$, it is unlikely that very high mass ratio,  high gas fraction mergers would be
detected by $G-M_{20}$. 

Our results here confirm the our conclusion in Paper 1 that
high gas-fraction mergers experience high asymmetries for longer durations than moderate gas-fraction mergers.  
In Figure \ref{tmorphgf0} we also plot the Sbc-Sbc prograde-prograde simulation time-scales from Paper 1. 
The Sbc-Sbc simulation has similar $f_{gas}$ to the G3gf2G3gf2 simulation but shows high asymmetry for $\sim$ 750 Myr shorter duration.
The difference between the Sbc-Sbc and G3gf2G3gf2 merger asymmetry time-scales suggests that gas scalelength and/or  dark matter concentration  may play secondary roles. 

We also compute the time-scales for detecting each merger as a close pair with projected separations $5 < R_{proj} < 20$, $10 < R_{proj} < 30$, 
$10 < R_{proj} < 50$, and $10 < R_{proj} < 100$ kpc $h^{-1}$ (Table 4) . 
The unequal-mass mergers show no dependence of the close pair time-scale on gas fraction.  However, the equal-mass mergers have
shorter close pair time-scales at higher gas fractions.   At small separations $< 30$ kpc $h^{-1}$,  this is because the higher gas fraction discs 
are less easily deblended. 
 
\section{Summary and Implications}
We have analysed the $g$-band quantitative morphologies and projected pair separations of a series of {\sc GADGET/SUNRISE}
simulations of disc galaxy mergers with increasing gas fractions.  These merger simulations span a range in baryonic mass 
ratio from 1:1 to 9:1,  and in primary galaxy baryonic gas-fraction $f_{gas}$ from 19\% to 53\%.  We determine the observability time-scales for 
identifying these simulated mergers using quantitative morphology classifications in  $G-M_{20}$, $G-A$, and $A$ and as close pairs with 
projected separations $5 < R_{proj} < 20$, $10 < R_{proj} < 30$, $10 < R_{proj} < 50$, and $10 < R_{proj} < 100$ kpc $h^{-1}$.  Our main conclusions
are as follows: 

$\bullet$ Gas-fraction has a strong effect on the quantitative morphologies of the merger simulations.  Both minor and major merger 
simulations with f$_{gas}$ $\ge$ 39\% show higher $A$, higher $M_{20}$, and lower $G$ values throughout the course of the merger. 
The high gas fractions produce more disc star-formation and therefore
higher $A$ and $M_{20}$ values for the initial disc galaxies.   During the course of a high gas-fraction merger, $A$ remains elevated after the 
first pass and after the final merger because of strong tidal features and residual star-formation.  $G$, on the other hand, is suppressed because of increased
dust extinction of the central nuclei. 

$\bullet$ The time-scale for detecting a galaxy merger with high asymmetry is a strong function of gas fraction.  
The asymmetry time-scale for a major merger with baryonic mass ratio $\leq$ 3:1 increases from $\sim$ 230 Myr for f$_{gas} = $ 19\%   
to $\sim$ 500$-$700 Myr for f$_{gas} = $ 39\% to $\sim$ 1$-$1.4 Gyr for f$_{gas} = $ 53\%.    While the 9:1 minor merger with f$_{gas} =$ 19\%  
does not produce asymmetries large enough to be classified as a merger,  the higher f$_{gas}$ minor mergers have asymmetry time-scales 
similar to the major mergers.  Therefore, at high $f_{gas}$,  asymmetry is equally likely to detect both minor and major mergers, but at low and moderate $f_{gas}$
asymmetry will primarily find major mergers. 

$\bullet$ The time-scale for detecting a galaxy merger via $G-M_{20}$ is not a strong function of the merger gas fraction, and
is weakly anti-correlated with $f_{gas}$ for minor mergers.  While the high $f_{gas}$
simulations produce more stars at large radii and increase $M_{20}$,  they have more dust obscuration for the central nuclei and
therefore lower $G$ values.  These two effects result in more mergers detected at the first pass and 
fewer mergers detected at the final merger by $G-M_{20}$,  but leave the total $G-M_{20}$ detection time-scale relatively unchanged ($\sim 250$ Myr). 

$\bullet$ The close pair time-scales do not change with increased gas fraction for the unequal-mass mergers.  Equal-mass mergers of high gas fraction discs are more
difficult to deblend at projected separations $<$ 30 kpc $h^{-1}$,  resulting in shorter projected separation time-scales. 

Gas-rich mergers are expected to be more common at redshifts $\ge 1$ than at present-day.  The fraction of galaxies in  blue galaxy $-$ blue galaxy pairs increases
strongly from $z \sim 0$ to $z \sim 1$ (Lin et al. 2008, Bundy et al. 2009, Ravel et al. 2009).  In addition,  the typical gas-fraction of galaxies may also evolve strongly with redshift.  Cold gas in $z \ge 1$ disc galaxies has recently been observed (Daddi et al. 2009; Tacconi et al. 2009),  and suggest gas-fractions $\ge$ 40\%,  significantly
higher than local discs.  These galaxies also have higher star-formation rates per unit mass (Noeske et al. 2007) 
and similar or smaller effective radii (Barden et al. 2005; Melbourne et al. 2007)  suggestive of higher gas densities than local disc galaxies.  

Many studies have found strong evolution in the fraction of bright or massive galaxies with high asymmetries from $z \sim 0$ to $z \ge 2$ 
(e.g. Abraham et al. 1996; Conselice et al. 2003, 2008;  Cassata et al. 2005; Lopez-Sanjuan et al. 2009; but see Shi et al. 2009).
Similar trends have also been found using visual classification of strongly-disturbed merger candidates and tidal tails 
(e.g. Brinchmann et al. 1998, Le F\'{e}vre et al. 2000,  but also see Jogee et al. 2009).  On the other hand, 
spectroscopically-confirmed pairs and $G-M_{20}$ techniques have found weak evolution in the 
fraction of galaxy merger candidates (Lin et al. 2004, Bundy et al. 2004, Lotz et al. 2008a, de Ravel et al. 2009).  

Our simulations predict that galaxy mergers will exhibit high asymmetries for longer periods of time if they have high gas-fractions.  
Asymmetric galaxies are also likely to be visually classified as mergers, therefore the time-scales for visually classifying a merger will also
scale with gas fraction.  This implies that much of the observed evolution in the asymmetric galaxy mergers may be the result of strong evolution 
in the typical gas-fraction of mergers, rather than the 
global merger rate.   In other words,  we find more high-redshift asymmetric galaxies not because mergers occur more frequently but because they are more likely 
to be gas-rich and/or they have higher gas-fractions than local mergers. 
On the other hand, the $G-M_{20}$ and close pair time-scales are expected to be independent of gas-fraction and therefore not affected by
by any evolution in the typical merger gas fraction.    This is consistent with weak evolution observed in the fraction of $G-M_{20}$ mergers and close pairs.

The merger simulations and observability time-scales presented here and in Papers 1 and 2 represent a dramatic step forward in our ability to interpret observations
of galaxy mergers.   With an understanding of how gas-fraction and merger mass ratio affect the detection of galaxy mergers, we can now determine a consistent galaxy merger
rate for each of different methods of identifying galaxy mergers.  We can also use the different behaviours of $G-M_{20}$, asymmetry, and close pairs to probe the 
evolution of mergers as a function of mass ratio and gas content. For a given merger, it is difficult to determine its mass ratio or gas-fraction from its morphology alone.  
 But by using the relative numbers of $G-M_{20}$ mergers, close pairs, and asymmetric objects, combined their observability time-scales, it will be possible to determine
 how the minor v. major and gas-rich v. gas-poor merger rates evolve with time.  We will explore these possibilities in more detail in our next paper (Paper 4).  
In the next few years,  the refurbished $HST$ with {\em Wide Field Camera 3} will 
certainly identify large samples of $z > 1$ morphologically disturbed galaxies.  With the commissioning of 
the {\em Atacama Large Millimeter Array},  it will be possible to determine if these merger candidates and other $z > 1$ galaxies have cold gas fractions 
substantially higher than local galaxies. 

JML acknowledges support from the NOAO Leo Goldberg Fellowship, NASA grants NAG5-11513 and HST-AR-9998, and 
would like to thank P. Madau for support for this project. 
PJ was supported by programs HST-AR-10678, HST-AR-10958, and HST-AR-11958 provided by NASA through grants from the Space Telescope Science
Institute, which is operated by the Association of Universities for Research in Astronomy, Incorporated, 
under NASA contract NAS5-26555, and  by the Spitzer Space Telescope Theoretical Research Program, through a contract 
issued by the Jet Propulsion Laboratory, California Institute of Technology under a contract with NASA.
TJC was supported by a grant from the W.M. Keck Foundation.  JRP was supported by NASA ATP grant NNX07AG94G. 

This research used computational resources of the
NASA Advanced Supercomputing Division (NAS) and the National Energy
Research Scientific Computing Center (NERSC), which is supported by
the Office of Science of the U.S. Department of Energy.

\end{document}